# Crystalline $C_{60}$ fulleride with metal inside

Ayano Nakagawa[1], Makiko Nishino[1], Hiroyuki Niwa[1], Katsuma Ishino[1], Zhiyong Wang[1], Haruka Omachi[1], Ko Furukawa[2], Takahisa Yamaguchi[3], Tatsuhisa Kato[3], Shunji Bandow[4], Jeremy Rio[5], Chris Ewels[5], Shinobu Aoyagi[6] & Hisanori Shinohara[1]*

[1]Department of Chemistry and Institute for Advanced Research, Nagoya University, Nagoya 464-8602, Japan.

[2]Center for Coordination of Research Facilities, Institute for Research Promotion, Niigata University, Niigata 950-2181, Japan.

[3]Graduate School of Human and Environmental Sciences, Kyoto University, Sakyo-ku, Kyoto 606-8501, Japan.

[4]Faculty of Science and Technology, Department of Applied Chemistry, Meijo University, Nagoya 468-8502, Japan.

[5]Institut des Materiaux Jean Rouxel (IMN), Université de Nantes, CNRS UMR6502, BP3229, 44322 Nantes, France

[6]Department of Information and Basic Science, Nagoya City University, Nagoya 467-8501, Japan.

Endohedral metallofullerenes have been extensively studied, since the first experimental observation of La@$C_{60}$ in a laser-vaporized supersonic beam in 1985. However, all of these studies have been carried out on metallofullerenes larger than $C_{60}$ such as (metal)@$C_{82}$, and there are no reported purified $C_{60}$-based metallofullerenes except for [Li@$C_{60}$]$^+$(SbCl$_6$)$^-$ salt. Pure (metal)@$C_{60}$ has not been obtained because of their extremely high chemical reactivity. We report here the first isolation, structural determination and electromagnetic properties of crystalline $C_{60}$-based metallofullerenes, Gd@$C_{60}$(CF$_3$)$_5$ and La@$C_{60}$(CF$_3$)$_5$. Synchrotron X-ray single-crystal diffraction reveals that La and Gd atoms are indeed encapsulated in the $I_h$-$C_{60}$ fullerene. The HOMO-LUMO gaps of Gd@$C_{60}$ and La@$C_{60}$ are significantly widened by an order of magnitude with addition of CF$_3$- groups. Magnetic measurements show the presne of a weak antiferromagnetic coupling in Gd@$C_{60}$(CF$_3$)$_3$ crystals at low temperatures. Gd@$C_{60}$ and La@$C_{60}$ might exhibit superconductivity as the electronic structures resemble those of superconducting alkali-doped $C_{60}$ fullerides.

$C_{60}$ fullerene is the most abundant fullerene among the so-called fullerene family[1-4]. However, $C_{60}$-based endohedral metallofullerenes (i.e., fullerenes with metal atom(s) encapsulated), referred to hereafter as $M@C_{60}$, have not been obtained in macroscopic and pure form to date. All of the previous studies on metallofullerenes reported during the past 25 years are based on higher fullerenes than $C_{60}$ such as $C_{80}$, $C_{82}$ and $C_{84}$[2], even though the presence of $M@C_{60}$ has been confirmed by mass spectrometry of supersonic cluster beams in the gas phase[3] as well as laser-vaporized processed soot[4]. Theoretical calculations also suggested the stability and possible structure of $M@C_{60}$[5-7]. The only exception is $[Li@C_{60}]^+(SbCl_6)^-$ salt, which was produced by $Li^+$ ion bombardment with $C_{60}$ onto a target plate in vacuum[8,9]. Neutral $Li^+@C_{60}^{\bullet-}$ radicals have been obtained by the electrochemical reduction of cationic $Li^+@C_{60}$; however, that has a dimerized form in the crystal[10]. $M@C_{60}$ have not been extracted by conventional fullerene solvents, preventing preparation of $M@C_{60}$ as macroscopic and pure materials.

The inability to extract and purify $M@C_{60}$ is due to their very small HOMO-LUMO gaps predicted, which lead to high chemical reactivity towards surrounding soot, air, moisture and various organic solvents. There have been, however, several attempts[11-14] to try and extract and purify $M@C_{60}$ by, for example, vacuum sublimation from soot followed by liquid chromatographic separation[12,13]. Unfortunately, none of these efforts led to the final macroscopic preparation of pure and isolated $M@C_{60}$. For this reason, $M@C_{60}$ are often referred to as the "missing metallofullerenes"[2,15,16].

One of the main reasons for the long-standing interest in $M@C_{60}$ is due to their expected novel electronic and magnetic behavior at low temperatures such as superconductivity, because the electronic band structures near the Fermi levels of $M^{3+}@C_{60}^{3-}$ solids resemble those of superconducting alkali-doped $C_{60}$ fullerides such as $(K_3)^{3+}C_{60}^{3-}$[17]. If this is the case, the presence of heavy atoms like lanthanoid elements inside $C_{60}$ might enhance the superconducting transition temperatures[18]. Here, we report the first macroscopic preparation of purified (>99.9 %) $Gd@C_{60}$ and $La@C_{60}$ as trifluoromethylated forms of $Gd@C_{60}(CF_3)_{3,5}$ and $La@C_{60}(CF_3)_{3,5}$, respectively, and their structural determination, electronic and magnetic properties.

**Results**

**Enhanced stabilization of $Gd@C_{60}$ and $La@C_{60}$.** To stabilize $Gd@C_{60}$ and $La@C_{60}$, we have employed *in situ* trifluoromethylation during the arc-discharge synthesis of the metallofullerenes developed in the present laboratory (see Supplementary Information and Fig. S1)[10,18]. For example, the *in situ* trifluoromethylation of $Gd@C_{60}$ generates $Gd@C_{60}(CF_3)_n$ (n=1-6) together with other sizes of trifluoromethylated Gd metallofullerenes, $Gd@C_{2m}(CF_3)_n$ (2m ≥ 70) (see Fig. S2). The high-temperature arc-discharge induces evaporation of polytetrafluoroethylene (PTFE) placed near the arc zone to produce $CF_3$ radicals[10]. We found trifluoromethyl-derivatized metallofullerenes, $Gd@C_{60}(CF_3)_n$ and $La@C_{60}(CF_3)_n$ (n=3,5), were formed efficiently. Trifluoromethyl derivatives of $Gd@C_{60}$ and $La@C_{60}$ are totally soluble and stable in toluene and carbon disulfide, which enabled us to perform HPLC purification and thus subsequent characterization. As discussed in later sections, the stability of these trifluoromethyl derivatives, $Gd@C_{60}(CF_3)_{3,5}$ and $La@C_{60}(CF_3)_{3,5}$, can be attributed to their closed-shell electronic structures, leading to wider HOMOng to gaps than those of pristine $Gd@C_{60}$ and $La@C_{60}$, respectively. With the current separation/isolation protocol (see Fig. S3), one may, for example, obtain ca. 1.0-2.0 mg of purified (>99.9 %) $Gd@C_{60}(CF_3)_5$ within 24 hours from the initial $o$-xylene extraction of the soot containing $Gd@C_{2m}(CF_3)_n$ obtained by ten arc-discharge syntheses (see Figs. S4 . is). The absorption onsets of the UV-Vis-NIR absorption spectra of $CS_2$ solution of $Gd@C_{60}(CF_3)_5$ and $La@C_{60}(CF_3)_5$ suggest the presence of enlarged HOMO-LUMO gaps of ca.1.2 eV comparable to that of $C_{60}$ (1.6 eV) (see Fig. S7). Interestingly, $Gd@C_{60}(CF_3)$ and $Gd@C_{60}(CF_3)_2$ have not been solvent-extracted because of their much smaller HOMO-LUMO gaps even though mass spectrometric analysis indicates the presence of these metallofullerenes in raw soot.

Because of the very high reactivity of $M@C_{60}$, it has not been self-evident that the metal is encapsulated in the conventional truncated icosahedral $I_h$-$C_{60}$. The $C_{60}$ might have a so-called non-IPR (isolated pentagon rule) structure[19,20], where two (or three) pentagons are fused with each other, since non-IPR fullerenes have frequently been observed in metallofullerenes during the past decade[2]. To answer this question, the molecular and crystal structures of $Gd@C_{60}(CF_3)_5$ (structural isomers I and II) and $La@C_{60}(CF_3)_5$ (I) have been determined by synchrotron radiation single-crystal X-ray diffraction at SPring-8 (see Supplementary Information).

**Synchrotron single-crystal X-ray diffraction measurements.** The X-ray results clearly show that the $C_{60}$ cage structures of $Gd@C_{60}(CF_3)_5$ (I, II) and $La@C_{60}(CF_3)_5$ (I) are very similar to that of the conventional empty $C_{60}(I_h)$. The five $CF_3$ groups of isomer I are attached to carbon atoms numbered as 6, 9, 12, 15 and 53 in a Schlegel diagram shown in Fig. 1a. In isomer II the five $CF_3$ groups are attached to carbon atoms numbered as 6, 9, 12, 15 and 36. In both isomers, four of the five $CF_3$ groups are attached to carbon atoms 6, 9, 12 and 15 bonded to carbon atoms 5, 1, 2 and 3 on a pentagon. Figures 1b,c and d,e show the molecular structures of $Gd@C_{60}(CF_3)_5$ (I) and (II), respectively. Very interestingly, the remaining $CF_3$ group is attached not to carbon atom 18 but to carbon atom 53 (isomer I) or 36 (isomer II) located far from the 1-2-3-4-5 pentagon. This asymmetric attachment of five $CF_3$ groups results in $C_1$ symmetry. If the 5$^{th}$ $CF_3$ group had been attached to carbon atom 18 bonded to carbon atom 4 on the pentagon, the molecule would have had five-fold symmetry. Instead of the 5$^{th}$ CF$_3$ group, a Gd (La) atom inside the C$_{60}$ cage is located near carbon atoms 18 and 4. The distances between the Gd atom and carbon atoms 18 and 4 are 2.351(8) and 2.387(8) Å in isomer I and 2.377(8) and 2.390(11) Å in isomer II, respectively.

The selective location of the Gd (La) atom is explained in terms of the elongation of C–C bond lengths by the attachment of CF$_3$ groups. The conventional C$_{60}$($I_h$) molecule consists of two kinds of C–C bonds, i.e., short 6 : 6 bonds fusing two hexagons with a bond length of 1.39 Å and longer 6 : 5 bonds fusing a hexagon and a pentagon with a bond length of 1.45 Å. As expected, the C–C bonds around the carbon atoms with CF$_3$ groups attached have sp$^3$ character, resulting in the elongation of the bonds. The bond lengths between 6-5, 9-1, 12-2 and 15-3 carbon atoms are ~1.51 Å in both of the isomers. The bond lengths between 18-4 atoms are 1.457(13) Å and 1.469(8) Å in Gd@C$_{60}$(CF$_3$)$_5$ (I) and (II), respectively, which are much longer than the normal 6 : 6 bond length (1.39 Å). The Gd is interacting strongly with the p$_z$-orbitals of atoms 18 and 4, resulting in the breaking of the $\pi$-bond and converts it to a single bond, hence the length of 1.457 Å which is a classical single-bond length. At the same time, there is a general attraction for the Gd towards the area of the C$_{60}$ surface that has been functionalized, hence this location beneath two carbon atoms rather than staying at the center of the neighbouring hexagon.

Gd@C$_{60}$(CF$_3$)$_5$ (I) and La@C$_{60}$(CF$_3$)$_5$ (I) form pure crystals that contain neither solvent nor ligand molecules. Figures 1f and g show the molecular packing of the Gd@C$_{60}$(CF$_3$)$_5$ (I) crystal. The centrosymmetric crystal with a space group of $P2_1/c$ contains two chiral isomers of Gd@C$_{60}$(CF$_3$)$_5$ (I) with the same number of molecules. Most of the endohedral metallofullerene crystals so far reported contain either solvent or ligand molecules.[2] Gd@C$_{60}$(CF$_3$)$_5$ (II) has been obtained only as a co-crystal with Ni(OEP) (OEP: octaethylporphyrin). The crystals of Gd@C$_{60}$(CF$_3$)$_5$ (I) and La@C$_{60}$(CF$_3$)$_5$ (I) are the first example of pure crystals of M@C$_{60}$ derivative. The crystal structure determined possesses a pseudo-hexagonal close-packed layer of the molecules in the *bc*-plane (Fig. 1f) and stacking of the layers along the *a*-axis (Fig. 1g). The stacking of the molecules is caused by intermolecular steric and electrostatic interactions. The molecule has a turtle-like structure (Figs. 1h and i), i.e., "a shell" of the C$_{60}$, "a head" of the isolated CF$_3$ group, "four legs" of the four CF$_3$ groups, and "a heart" of the encapsulated Gd (La) atom. The molecule has an electric dipole moment (*P*) because of the electron-withdrawing CF$_3$ groups asymmetrically attached to the C$_{60}$ cage and the presence of the off-centered Gd (La) atom inside. The stacking of the turtle-like molecules with an electric dipole moment can provide an electrostatically stable close-packed allay along the *a*-axis as shown in Fig. 1h (see Figs. S8 – S9). The detailed crystallographic data are summarized in Supplementary Information and Tables S1-S6.

We then carried out $^{19}$F-NMR measurements of La@C$_{60}$(CF$_3$)$_n$ (see Supplementary Information and Fig. S12). The spectra of the two La@C$_{60}$(CF$_3$)$_5$ isomers consist of five equal-intensity peaks, which are four quartets (or multiplets) due to the through-space coupling of $CF_3$ legs and the other singlet due to $CF_3$ head[21,22]. In contrast, the $^{19}F$ NMR spectrum of $La@C_{60}(CF_3)_3$ only showed a singlet peak at –75.5 ppm because the three trifluoromethyl groups are geometrically equivalent, indicating $C_3$ symmetry of $La@C_{60}(CF_3)_3$.

**Discussion**

To obtain information on the growth mechanism, structure and electronic properties of $Gd@C_{60}(CF_3)_5$, we perform spin unrestricted density functional (DFT) calculations under the local spin density approximation[23-25], applying relativistic pseudopotentials[26]. Wavefunctions are constructed from 38/90/28 independent Gaussian based functions for C/Gd/F, respectively, up to $l=3$ angular momentum, with an electron temperature of $k_BT=0.04$ eV for level occupation and 300 Ha cut-off plane-waves for the charge density. To explore the addition pathways for $CF_3$ functionalization of $Gd@C_{60}(CF_3)_n$, $n=0-5$, we assume that $CF_3$ adds to the most thermodynamically favored site and is then immobile, which is justified by a calculated high surface migration barrier of $CF_3$ ($n=1$) of 59 kcal/mol. Taking the most stable $n$th isomer as a starting point for $n+1$ stepwise addition allows us to reduce the number of isomers considered to a manageable several hundred. Where two stable isomers are closer in energy than 2.3 kcal/mol of each other, we have used both as starting points for subsequent addition, allowing us to trace bifurcations in the addition pathway.

The Gd atom in $Gd@C_{60}$ is initially covalently bound beneath hexagon (3-4-15-16-17-18) with 2.41Å Gd-C bond lengths. The most stable $CF_3$ addition pathway involves sequential addition to back bonds of the 1-2-3-4-5 pentagon, i.e., addition at sites 6, 9, 12 and 15 in that order shown in top line of Fig. 2a. Instead of completing the pentagon at site 18, which is excluded due to the presence of Gd below this last pentagon back bond with ~2.35Å Gd-C bond lengths, the fifth $CF_3$ group adds behind the neighbouring pentagon at sites 53 (isomer I) or 36 (isomer II). Adding a sixth $CF_3$ group is endothermic compared to ½$(CF_3)_2$, confirming that $n=5$ is the end point in the addition sequence in agreement with the experimental observation.

After $n=1$ addition at site 6 there is a second stable isomer for $n=2$, where Gd maintains its hexagonal site and $CF_3$ adds to site 36 (lower line of Fig. 2a). The pathway then bifurcates once again, the energetically favoured route continuing the 9, 12, 15 addition sequence to isomer II. However an alternative $n=3$ addition at site 13 results instead in a $C_3$ symmetry $Gd@C_{60}(CF_3)_3$ isomer. Further $CF_3$ addition to this is endothermic, showing this to be a third stable end-point in the addition sequence. This mapping process thus shows remarkable agreement with the current X-ray and $^{19}F$-NMR experimental results, successfully predicting the three observed $Gd@C_{60}(CF_3)_n$ species, i.e., $Gd@C_{60}(CF_3)_5$ (I), $Gd@C_{60}(CF_3)_5$ (II) and $Gd@C_{60}(CF_3)_3$.

These three stable isomers have multiplicity 7/2, associated with the Gd atom. They are closed shell systems (unlike many of the less stable isomers), restoring the gap between highest occupied molecular orbital (HOMO) and lowest unoccupied molecular orbital (LUMO) to ~2/3 that of $C_{60}$ (Fig. 2b) with the largest gap for $Gd@C_{60}(CF_3)_3$. The unpaired Gd *f*-states are deep lying with the HOMO and LUMO localized primarily over the fullerene cage with minor $CF_3$ character (Fig. 2c). Gd plays an interesting dual role when selecting $CF_3$ addition sites, acting as donor and localizing surface charge, while also affecting surface chemistry through strong hybridization with the carbon 2p states. This suggests different addition sequences are likely for non-covalently bound endohedral +3 oxidation species. As seen above, the agreement between the X-ray and the theoretical results for the presence of the stable isomers of $Gd@C_{60}(CF_3)_{3,5}$ strongly suggests that $CF_3$ groups are sequentially added to $Gd@C_{60}$ during the arc-discharge synthesis. The energy levels, HOMO-LUMO and spin-up and –down charge density for $Gd@C_{60}(CF_3)_3$ and $Gd@C_{60}(CF_3)_5$ (I,II) are shown in Supplementary Information Fig. S13 and S14, respectively.

The magnetic properties of $Gd@C_{60}(CF_3)_3$ metallofullerene was investigated by SQUID measurements (see Supplementary Information). The HOMO-LUMO gap of $Gd@C_{60}(CF_3)_3$ (1.4 eV) is larger than that of $Gd@C_{60}(CF_3)_5$ (1.2 eV) suitable for solvent-extraction and HPLC isolation. Magnetization curves of solid $Gd@C_{60}(CF_3)_3$ indicate no hysteresis as shown in Fig. 3, where the ordinate is normalized by experimentally determined saturation magnetization $M_s$ of 41,108 emu·G/mol. Except for the magnetization curve at 2 K, other data points are on the same curvature trace. When magnetic moments thermally fluctuate without strong interaction between the magnetic moments, magnetization curves are fitted by the following Brillouin function $B_J(x)$

$$M = M_s B_J(x)$$

$$M_s = NgJ\mu_B$$
$$B_J(x) = \frac{2J+1}{2J}\coth\left(\frac{2J+1}{2J}x\right) - \frac{1}{2J}\coth\left(\frac{1}{2J}x\right) \qquad (1)$$
$$x = \frac{gJ\mu_B H}{k_B T}$$

where $N$ the spin concentration, $g$ the g-factor, $J$ the total quantum number, $\mu_B$ the Bohr magneton and $k_B$ the Boltzmann constant.  From the curve fitting, we found that the total quantum number $J = 7/2$ is needed to explain the experimental results by setting $g = 2$.  This gives an effective magnetic moment $\mu_{eff}$ ($= g\sqrt{J(J+1)}\,\mu_B$) of 7.94$\mu_B$ per $Gd@C_{60}(CF_3)_3$ molecule.  This experimental analysis supports a trivalent $Gd^{3+}$ state by transferring 3 electrons to the $C_{60}$ cage ($Gd^{3+}@C_{60}^{3-}$) and these electrons are then used to make chemical bonds with three $CF_3$ on $C_{60}$. Mulliken charge analysis shows that most of the transferred charge is indeed spread around the $C_{60}$ cage beneath the $CF_3$ groups, on the central pentagon, and on the neighbours of the Gd. The magnetic moment is thus essentially located on the encaged $Gd^{3+}$ ion ($4f^7$, $J = 7/2$ with orbital quantum number $L =$

0, *i.e.*, $Gd^{3+}$ has spin $7\mu_B$) in agreement with the theoretical result shown in Fig. 2b. According to the previous studies on $M^{3+}@C_{82}^{3-}$ (M = Gd, Dy, Ho, Er) metallofullerenes[27-31], the magnetic moments observed are significantly small compared to the expected moments of free $M^{3+}$ ions.

This reduction of the magnetic moments was theoretically explained by antiferromagnetic coupling between the trivalent endohedral metals and free spins on the carbon cage[32]. In the present Gd (La)-encapsulating $C_{60}$ fullerides, since free spins on $C_{60}$ are very small by forming the chemical bonds with $CF_3$ on the cage, this guarantees an almost ideal magnetic moment of $Gd^{3+}$ ion. Theoretically, a trace counter-spin can be seen localised on the $C_{60}$ cage between the Gd and $CF_3$ groups at 0 K (Fig 2c), presumably too weak to be detectable by SQUID but partially responsible for the weak antiferromagnetic coupling observed at low temperature. This was also confirmed by the analysis of temperature dependence of molar magnetic susceptibility as explained in Supplementary Information.

To further investigate the electronic structures of $Gd@C_{60}(CF_3)_5$ (I,II), we also performed electron spin resonance (ESR) measurements in $CS_2$ frozen solution (see Supplementary Information and Table. S7). The spectra recorded by an X-band CW ESR spectrometer at 4K and by a W-band at 20K are shown in the upper and lower panels of Fig. 4, respectively, where the observed spectra with fine structure in red are compared with simulated ones in blue[33,34]. The parameters shown in Table S7 were determined so that both of the observed spectra by the X- and W-band measurements should simultaneously be reproduced by the simulation. In both cases the spin quantum number of S=7/2 was obtained, consistent with the theoretical and magnetic measurement results described above. The zero-field splitting parameters, $D$=0.162 cm$^{-1}$ and $D$=0.193 cm$^{-1}$, were obtained for $Gd@C_{60}(CF_3)_5$ (I) and $Gd@C_{60}(CF_3)_5$ (II), respectively, comparable to those for S=7/2 states of $Gd@C_{82}$[33] and $Eu@C_{82}$[34]. The lift of the *x* and *y* degeneracy of $E$=0.012 cm$^{-1}$ for $Gd@C_{60}(CF_3)_5$ (II), which is bigger than that ($E$=0.000 cm$^{-1}$) of $Gd@C_{60}(CF_3)_5$ (I), suggests that the position of the $Gd^{3+}$ ions in the two isomers differs slightly, since the *E* parameters originate basically in the spin-orbit coupling of the $Gd^{3+}$ ions. Our calculations indeed indicate that the $Gd^{3+}$ ion of isomer (I) lies more symmetrically beneath the C–C bond (so more symmetrically with respect to the four $CF_3$ groups around the pentagon) than that of isomer (II). The observed ESR parameters are summarized in Supplementary Information Table S7.

In summary, the missing metallofullerenes, $Gd@C_{60}$ and $La@C_{60}$, in the trifluoromethylated forms of $Gd@C_{60}(CF_3)_{3,5}$ and $La@C_{60}(CF_3)_5$, have been obtained in pure bulk forms for the first time. *In-situ* trifluoromethylation widens the HOMO-LUMO gaps of $Gd@C_{60}$ and $La@C_{60}$, enabling isolation of the metallofullerenes and the subsequent single-crystal synchrotron X-ray diffraction. The crystals exhibit a pseudo-hexagonal close-packed layer of molecules in the *bc*-plane and stacking of the layers along the *a*-axis (a turtle-like stacked structure). The temperature dependence of the magnetic susceptibility of Gd@$C_{60}$(CF$_3$)$_3$ is well fitted by the Curie-Weiss law. The magnetic moment is located on the encaged Gd$^{3+}$ ion (4$f^{\,7}$, $J = 7/2$ with orbital quantum number $L = 0$), with very small magnetic moment on the cage in agreement with the theoretical results. The intact forms of Gd@$C_{60}$ and La@$C_{60}$ might exhibit superconductivity as the electronic structures resemble those of superconducting alkali-doped $C_{60}$ fullerides. The preparation and isolation of intact Gd@$C_{60}$ fulleride is now underway in the present laboratory.

## Methods

**Synthesis, purification and crystallization of Gd@$C_{60}$(CF$_3$)$_5$ (I), La@$C_{60}$(CF$_3$)$_5$ (I), Gd@$C_{60}$(CF$_3$)$_5$ (II) and Gd@$C_{60}$(CF$_3$)$_3$.** A cross-sectional view of the DC arc-discharge chamber is illustrated in Fig. S1, where polytetrafluoroethene (PTFE) rods (40 g) are placed near the discharge area. Graphite rods (100 g) impregnated with Gd (La) (0.8 mol %, Toyo Tanso Co.LTD) was used as the anode. A pure graphite rod (Toyo Tanso Co.LTD) was used as the cathode. Arc discharge was performed at a DC current of 500 A in a flowing He atmosphere with a pressure of 7-9 kPa. During arc-discharge because of the high-temperature around the arc zone, PTFE was decomposed and evaporated to produce CF$_3$ radicals. Normally, 50–70 g of raw soot was obtained per discharge. Gd-metallofullerenes and empty fullerenes were extracted from the raw soot with o-xylene.

**Extraction and separation of trifluoromethylated Gd-metallofullerenes from empty fullerenes by TiCl$_4$.** The rapid separation of the metallofullerenes from empty fullerenes was carried out by the TiCl$_4$ Lewis acid method developed in the present laboratory. To a 500 mL CS$_2$ solution of the crude mixture of Gd-fullerenes and empty fullerenes, ca. 5mL of TiCl$_4$ was added. Metallofullerenes were reacted immediately and insoluble complexes were precipitated out. After mixing for 5 minutes, the precipitate was collected on a PTFE membrane filter and washed with 10–20 mL of CS$_2$ to separate from the empty fullerenes solution. Deionized water was passed through the filter to decompose the complex of metallofullerene/TiCl$_4$, and then washed with acetone to eliminate extra water. Finally, CS$_2$ was passed through the filter to collect desired Gd-metallofullerenes as a solution as shown in Fig. S2.

**Multi-stage HPLC purification of Gd-metallofullerenes.** High-performance liquid chromatography (HPLC) purification was conducted by using a JAI (Japan Analytical Industry Co. LTD.) recycling preparative HPLC LC-9104HS. The overall separation and isolation scheme of Gd@$C_{60}$(CF$_3$)$_n$ is shown in Fig. S3. For the identification of metallofullerenes, mass spectrometric analysis was performed on Shimadzu MALDI-TOF-MS Spectrometer. Vis/NIR absorption spectra of metallofullerenes in $CS_2$ were recorded by a Jasco V-570 spectrophotometer. Three isomers of $Gd@C_{60}(CF_3)_n$ were isolated from the mixture by the multi-stage high-performance liquid chromatography (HPLC) method developed in the present laboratory. Two kinds of columns were used alternatively with toluene eluent for the isolation, i.e., Buckyprep column (20 mm diameter × 250 mm, Nacalai Tesque Inc.) and Buckyprep-M column (20 mm diameter × 250 mm, Nacalai Tesque Inc.). The initial (the first-stage) HPLC purification was performed with Buckyprep-M. $Gd@C_{60}(CF_3)_n$ (n=3,5) was obtained in fraction 1. The HPLC chromatograms are shown Fig. S4. Identification and isolation of the metallofullerenes were checked by MALDI mass spectroscopy as shown in Figs. S5 and S6. UV-Vis-NIR absorption spectra of the metallofullerenes exhibit characteristic features and also provide an estimate on their HOMO-LUMO gaps judging from the absorption onsets. The spectral features between Gd- and La-metallofullerenes are almost the same with each isomers as shown in Fig. S7.

**Synchrotron single-crystal X-ray structure analysis.** Single-crystal X-ray diffraction (XRD) measurements of $Gd@C_{60}(CF_3)_5$ (I), $La@C_{60}(CF_3)_5$ (I), $Gd@C_{60}(CF_3)_5$ (II) and $Gd@C_{60}(CF_3)_3$ were performed at SPring-8 BL02B1. The crystallographic data and various bond lengths for $Gd@C_{60}(CF_3)_5$ (I,II), $Gd@C_{60}(CF_3)_3$ and $La@C_{60}(CF_3)_5$ (I) are summarized in Table S1-S6. The crystal structures for $Gd@C_{60}(CF_3)_5$ (I), $La@C_{60}(CF_3)_5$ (I) and $Gd@C_{60}(CF_3)_5$ (II) are shown in Figs. S8, S9 and S10, respectively. The charge density surface of $Gd@C_{60}(CF_3)_3$ is shown in Fig. S11. $Gd@C_{60}(CF_3)_5$ (I) and $La@C_{60}(CF_3)_5$ (I): Single crystals of $Gd@C_{60}(CF_3)_5$ (I) and $La@C_{60}(CF_3)_5$ (I) were obtained from $CS_2$ solution by vapor diffusion. Results of the XRD measurement were summarized in Table S1. The crystal structures were determined by using *SIR* and *SHELX* with good reliable factors. $Gd@C_{60}(CF_3)_5$ (I) and $La@C_{60}(CF_3)_5$ (I) have similar crystal structures. The centrosymmetric monoclinic crystals consist of the same number of two chiral isomers of the molecule. The unit cell contains two right-handed isomers and two left-handed isomers as shown in Figures S8a and S9a. An independent molecule in the asymmetric unit has a disordered structure in which two chiral isomers overlap with the ratios of 0.8 and 0.2 as shown in Figures S8b and S9b. 95 C−C, 15 C−F, and 4 C−Gd (La) bond lengths of the major part of the disordered $Gd@C_{60}(CF_3)_5$ (I) and $La@C_{60}(CF_3)_5$ (I) are listed in Table S2 and S3, respectively. 30 short 6 : 6 bonds fusing two hexagons and 60 long 6 : 5 bonds fusing a hexagon and a pentagon on the $C_{60}$ cage are separately shown. The CIF deposition numbers at the Cambridge Crystallographic Data Centre (CCDC) are 1587428 for $Gd@C_{60}(CF_3)_5$ (I) and 1587430 for $La@C_{60}(CF_3)_5$ (I). $Gd@C_{60}(CF_3)_5$ (II): Single crystals of $Gd@C_{60}(CF_3)_5$ (II) were obtained from $CS_2$ solution as co-crystals with Ni(OEP) (OEP: octaethylpolphyrin) by vapor diffusion. Results of the XRD measurement were summarized in Table S4. The crystal structure was determined by using *SIR* and *SHELX* with a good reliable factor. The centrosymmetric triclinic crystals consist of the same number of two chiral isomers of Gd@C$_{60}$(CF$_3$)$_5$ (II). The unit cell contains a right-handed isomer, a left-handed isomer, three Ni(OEP) and half toluene molecules as shown in Figure S10a. An independent molecule in the asymmetric unit has a disordered structure in which two chiral isomers overlap with the ratios of 0.8 and 0.2 as shown in Figure S10b. 95 C−C, 15 C−F, and 4 C−Gd bond lengths of the major part of the disordered Gd@C$_{60}$(CF$_3$)$_5$ (II) are listed in Table S5. 30 short 6 : 6 bonds fusing two hexagons and 60 long 6 : 5 bonds fusing a hexagon and a pentagon on the C$_{60}$ cage are separately shown. The CIF deposition number at CCDC is 1587429. Gd@C$_{60}$(CF$_3$)$_3$: Single crystals of Gd@C$_{60}$(CF$_3$)$_3$ were obtained from CS$_2$ solution by vapor diffusion. Results of the XRD measurement were summarized in Table S6. The tetragonal unit cell can contain 16 Gd@C$_{60}$(CF$_3$)$_3$ molecules. An expected molecular arrangement was obtained by charge flipping using *Superflip* and maximum entropy method using *ENIGMA* as shown in Figure S11. The figure shows an electron charge density surface obtained by maximum entropy method. A structure model with 16 double uniform shells of Gd@C$_{60}$ located at (1/8, 1/8, 0), (3/8, 1/8, 0), (1/8, 3/8, 0), (3/8, 3/8, 0), (7/8, 7/8, 0), (5/8, 7/8, 0), (7/8, 5/8, 0), (5/8, 5/8, 0), (1/8, 7/8, 1/2), (1/8, 5/8, 1/2), (3/8, 7/8, 1/2), (3/8, 5/8, 1/2), (7/8, 1/8, 1/2), (7/8, 3/8, 1/2), (5/8, 1/8, 1/2), and (5/8, 3/8, 1/2) was used in the analysis. Detailed structure of Gd@C$_{60}$(CF$_3$)$_3$ could not be determined due to a severe orientation disorder and lack of resolution. We also attempted to obtain co-crystals of Gd@C$_{60}$(CF$_3$)$_3$ with Ni(OEP) (OEP: octaethylpolphyrin); however, that have not been obtained.

**$^{19}$F-NMR measurements.** Nuclear magnetic resonance (NMR) spectra were recorded on a JEOL ECS-400 spectrometer; chemical shifts for $^{19}$F NMR (376 MHz, CDCl$_3$/CS$_2$) are expressed in parts per million (ppm) relative to C$_6$F$_6$ ($\delta$ = –164.9 ppm). Data are reported as follows: chemical shift, multiplicity (s = singlet, q = quartet, m = multiplet), coupling constant (Hz), and integration. La@C$_{60}$(CF$_3$)$_3$: $\delta$ –75.5 (s, 9F). La@C$_{60}$(CF$_3$)$_5$ (I): $\delta$ –79.4 (s, 3F), –76.0 (q, *J* = 10 Hz, 3F), –74.9 (q, *J* = 12 Hz, 3F), –67.4 (m, 3F), –66.9 (m, 3F). La@C$_{60}$(CF$_3$)$_5$ (II): $\delta$ –78.9 (s, 3F), –78.0 (q, *J* = 12 Hz, 3F), –73.6 (q, *J* = 10 Hz, 3F), –68.3 (m, 3F), –67.3 (m, 3F).

**SQUID magnetic measurement.** Sample de-solved in CS$_2$ was transferred into quartz tube and the solution was vaporized under argon flow. Then the sample in quartz tube was evacuated by using turbo molecular pump at 260 ºC for 12 hours and vacuum sealed. By this heat treatment, weight of Gd@C$_{60}$(CF$_3$)$_3$ sample was decreased by ~65 % (finally 0.71 mg of sample was loaded in the quartz tube), while no such weight change was observed for La@C$_{60}$(CF$_3$)$_5$ one. Former finding suggests that the CS$_2$ molecules were incorporated in a Gd@C$_{60}$(CF$_3$)$_3$ solid, forming Gd@C$_{60}$(CF$_3$)$_3$(CS$_2$)$_{6.4}$ just after vaporizing CS$_2$, and incorporated CS$_2$ molecules have been removed by heat treatment. On the other hand, La@C$_{60}$(CF$_3$)$_5$ solid did not incorporate any CS$_2$ molecule. The temperature dependence of molar magnetic susceptibility $\chi_m$ of Gd@C$_{60}$(CF$_3$)$_3$ is shown in the top panel of Fig. S15a, which is well fitted by the Curie-Weiss law represented by

$$\chi_m = \chi_c + \frac{N g^2 J(J+1) \mu_B^2}{3k_B(T-\Theta)} = \chi_c + \frac{N \mu_{eff}^2}{3k_B(T-\Theta)}, \quad (1)$$

with $J = 7/2$, $\Theta = -0.14$ and $N = 6.34 \times 10^{23}$ mol$^{-1}$ (almost the same as Avogadro's number), where $\chi_c$ is a temperature independent constant term and $\Theta$ the Weiss temperature. This also supports a $\mu_{eff}$ of 7.94$\mu_B$ for each Gd@C$_{60}$(CF$_3$)$_3$ molecule. The bottom panel of Fig. S15a indicates that the extrapolation of a slope of inverse magnetic susceptibility against temperature crosses at a temperature of $-0.14$ K, which suggests that a very small antiferromagnetic coupling exists between the magnetic moments on gadolinium ions. It is thus reasonable to observe a small decrease of $J$ at 2 K as seen in the bottom panel of Fig. 3. From the analyses of magnetization curves and temperature dependence of $\chi_m$, we obtain $J = 7/2$ and $N = 6.34 \times 10^{23}$ mol$^{-1}$. Using these parameters, we obtained an expected value for $M_s$ ($= NgJ\mu_B$) of 41,140 emu·G/mol, in good agreement with $M_s$ of 41,108 emu·G/mol obtained from the magnetization curve measurements. However, if each Gd@C$_{60}$(CF$_3$)$_3$ molecule has $J = 7/2$, $N$ should be equal to $N_A$, giving an ideal value for $M_s$ ($= N_A g J \mu_B$) of 39,064 emu·G/mol. We think that the difference between this ideal value and experimental $M_s$ originates in the measurement error of the sample weight loaded in the SQUID quartz tube, *i.e.*, an inaccuracy of ~0.04 mg with respect to 0.71 mg. The magnetization curve of La@C$_{60}$(CF$_3$)$_5$ solid shows a linear response with respect to the magnetic field down to 2 K, suggesting a small $\mu_{eff}$ if it exists. The temperature dependence of $\chi_m$ behaves Curie-like as shown in Fig. S15b, and is roughly fitted by

$$\chi_m = \chi_c + \frac{N_A \mu_{eff}^2}{3k_B T}, \quad (2)$$

with $\mu_{eff}$ of 0.35$\mu_B$ per La@C$_{60}$(CF$_3$)$_5$ molecule, where $N_A$ is Avogadro's number. Hence, 3 electrons of encaged La are transferred to the cage (La$^{3+}$@C$_{60}$$^{3-}$) giving La$^{3+}$ (5$p^6$, $J = 0$) and these electrons are distributed to five CF$_3$ in order to make chemical bonds with C$_{60}$. This may cause a small $\mu_{eff}$ on the cage due to the difference between $\alpha$ and $\beta$ spin concentrations. The $\mu_{eff}$ of La@C$_{60}$(CF$_3$)$_5$ is almost the same with that of La@C$_{82}$. The valency of La of La@C$_{82}$ is also 3+, and the transferred electrons to the C$_{82}$ cage are uniformly diffused on the cage. We therefore think that a similar magnetic feature was observed for the present La@C$_{60}$ fulleride.

**Electron spin resonance (ESR) measurements.** The X- and W-band ESR measurements were performed using a Bruker E500 and a E680 spectrometer, respectively. The temperature was controlled by helium flow cryostat (Oxford Instruments model ITC500). The X-band ESR spectra were measured at 4.0K, and W-band ESR spectra were measured at 20K. The spectral simulation were performed by using MATLAB software package with EasySpin toolbox. The observed ESR parameters are shown in Table S7.

## Acknowledgments

We thank David Tomanek for valuable discussion. The experimental part of this work was financially supported by MEXT/JSPS KAKENHI Grant Numbers JP16H06350 and JP16H02248. The SQUID measurement part was financially supported by the Private University strategic research infrastructure support program of MEXT based on green innovation research at Meijo University, S1511021. The theory part was financially supported by Region Pays de la Loire "Paris Scientifiques 2017" Grant Number 09375 and CCIPL for computing resources. The synchrotron radiation experiments were performed at SPring-8 with the approval of the Japan Synchrotron Radiation Research Institute (JASRI) (Proposal No. 2017A1206 and 2017B1373).

## Author contributions

H.S. led the entire project and contributed to the experimental work including synthetic apparatus design of the present metallofullerenes. A.N., M.N., H.N., K.I., and Z.W. contributed to the synthesis, purification and crystal growth of the materials. H.O. and A.N. contributed to [19]F-NMR measurement. K.F., T.Y. and T.K. contributed to the ESR measurement. S.B. contributed to the SQUID magnetic measurement. S.A. contributed to the synchrotron X-ray diffraction measurement at SPring-8 and the data analyses. J.R. and C.E. contributed to the theoretical part of the work. H.S., T.K., S.B., S.A. and C.E. wrote the paper.

## Additional information

**Supplementary Information** accompanies this paper at ………………

**Competing interests**: The authors declare no competing financial interests.

**Reprints and permission** information is available online at …….

**Publisher's note**: Springer Nature remains neutral with regard to jurisdictional claims in published maps and institutional affiliations.

# Figure Captions

**Figure 1 | Structures of Gd@C$_{60}$(CF$_3$)$_5$ (I) and (II). a,** Schlegel diagram of C$_{60}$. Carbon atoms CF$_3$ attached to Gd@C$_{60}$(CF$_3$)$_5$ (I) and (II) are shown as red circles. **b** and **c,** Molecular structure of Gd@C$_{60}$(CF$_3$)$_5$ (I). **d** and **e,** Molecular structure of Gd@C$_{60}$(CF$_3$)$_5$ (II). Front views facing the bond between carbon atoms 4 and 18 (**b** and **d**) and side views (**c** and **e**) are shown. The thermal ellipsoids in **b-e** are drawn at 50% probability level. **f** and **g,** Molecular packing in the Gd@C$_{60}$(CF$_3$)$_5$ (I) crystal viewed along the *a*- and *c*-axis, respectively. Pseudo-hexagonal close-packed layers in the *bc*-plane are stacked along the *a*-axis. **h,** Steric and electrostatic stacking of the Gd@C$_{60}$(CF$_3$)$_5$ (I) molecules with an electric dipole moment (*P*) along the *a*-axis. The polar Gd@C$_{60}$(CF$_3$)$_5$ (I) molecules with a CF$_3$ head, four CF$_3$ legs and a Gd heart form a close-packed array similar to stacking turtles shown in **i**.

**Figure 2 | a. Schlegel projections (see Fig. 1a) showing calculated CF$_3$ addition pathways to Gd@C$_{60}$.** Previous addition sites marked in blue, new addition site in red. Pink shading indicates location of the encapsulated Gd atom. Numbers indicate energy release in kcal/mol according to the reaction Gd@C$_{60}$(CF$_3$)$_{n-1}$ + ½ (CF$_3$)$_2$ → Gd@C$_{60}$(CF$_3$)$_n$. All pathways are shown where isomers are within 2.3kcal/mol of the most stable. Fullerenes where next addition is endothermic have shaded backgrounds and represent stable end points of a reaction pathway. **b.** Calculated Kohn-Sham eigenvalues (eV) for (left) C$_{60}$, (centre) Gd@C$_{60}$, and (right) Gd@C$_{60}$(CF$_3$)$_3$, showing CF$_3$ functionalization results in a closed shell system, reopening the HOMO-LUMO gap to nearly that of C$_{60}$. **c** Spatial distribution of calculated Kohn-Sham eigenstates showing highest occupied molecular orbital (HOMO) and lowest unoccupied molecular orbital (LUMO) and their neighbouring states around the Fermi level for the most stable Gd@C$_{60}$(CF$_3$)$_5$(I) isomer. The states are delocalised across the fullerene cage with very limited CF$_3$ character.

**Figure 3 | Normalized magnetization curves of Gd@C$_{60}$(CF$_3$)$_3$.** Top panel shows whole trace of magnetization taken at 273, 100, 20, 10, 4.2 and 2 K, and bottom is an expansion around the origin of abscissa. $M_s$ value determined experimentally is 41108 emu·G/mol. Solid lines are the calculated magnetization curves using Brillouin function with $J$ = 8/2, 7/2 and 6/2. Red open circles are for the data taken at 2 K. Only this data set did not fit with the same curvature.

**Figure 4 | Confirmation of ESR parameters for Gd@C$_{60}$(CF$_3$)$_5$(I) and Gd@C$_{60}$(CF$_3$)$_5$(II). a, c.** ESR spectra (red lines) of Gd@C$_{60}$(CF$_3$)$_5$(I) and Gd@C$_{60}$(CF$_3$)$_5$(II) recorded by X-band (9GHz) spectrometer in CS$_2$ at 4K, **b,d.** W-band (90GHz) spectrometer in CS$_2$ at 20K. ESR parameters were determined so that both observed spectra in **a** and **b, c and d,** should be simultaneously reproduced by simulation (blue lines).

**Figures**

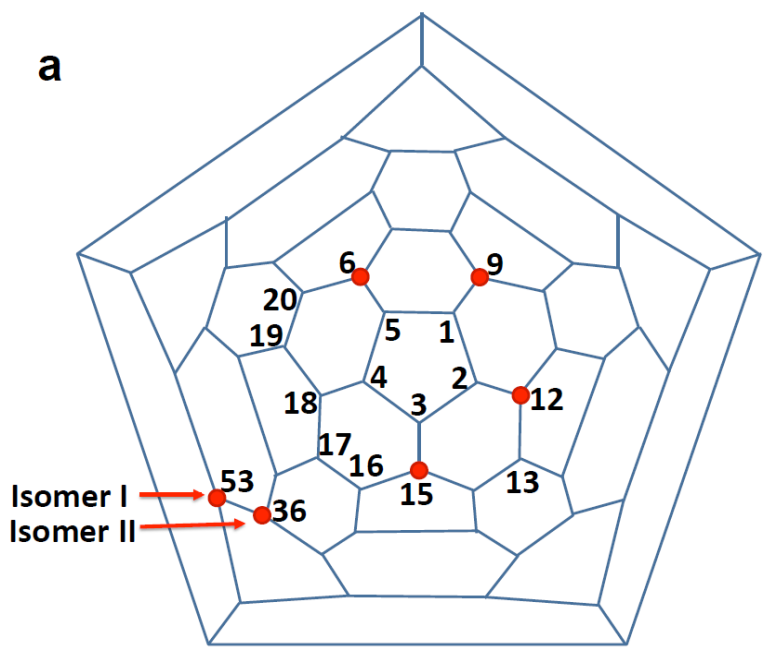

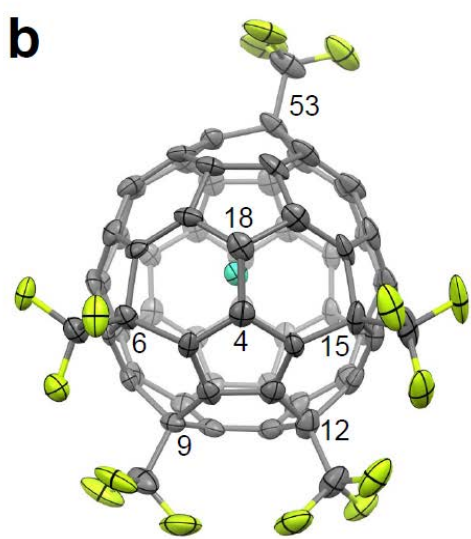

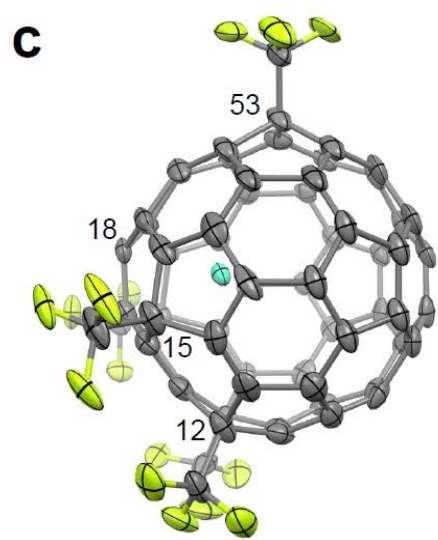

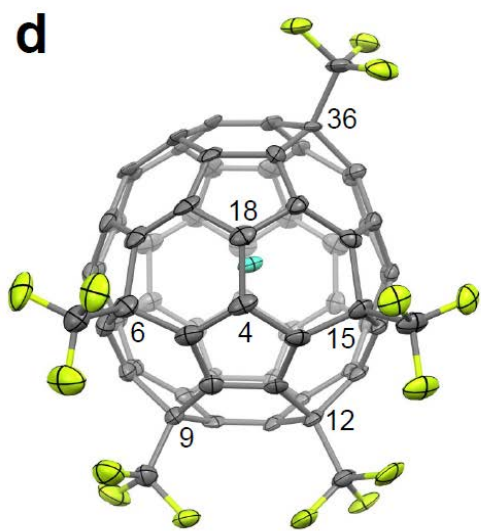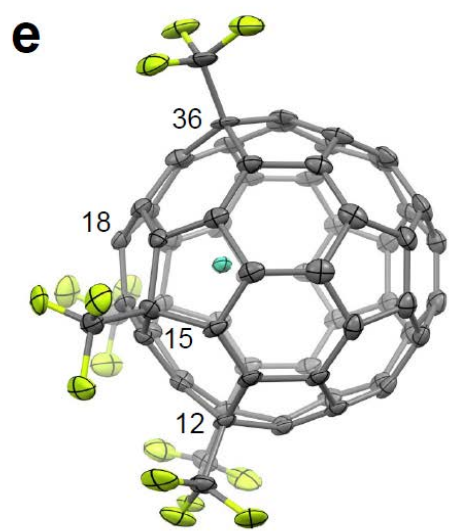

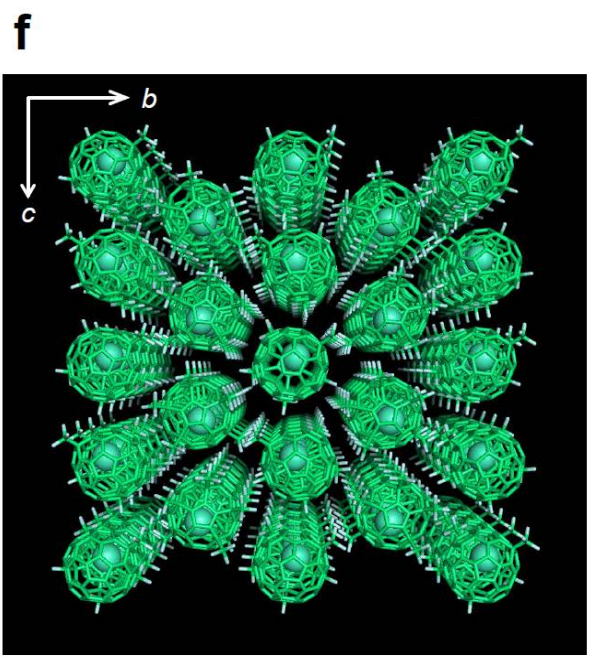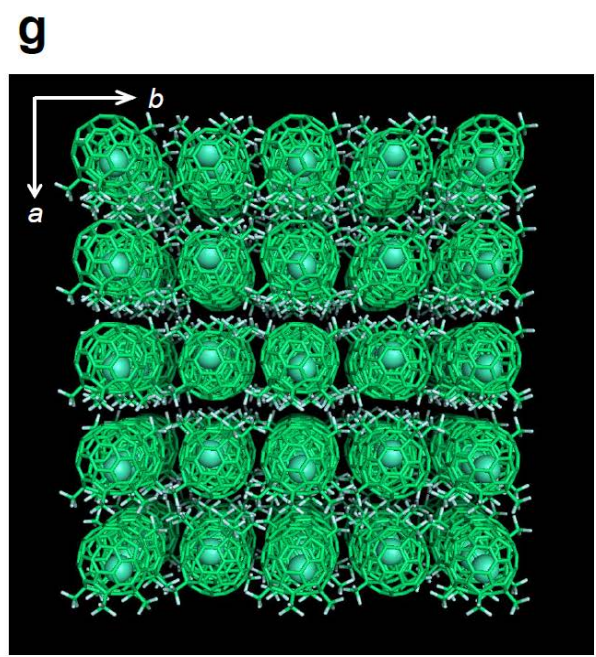

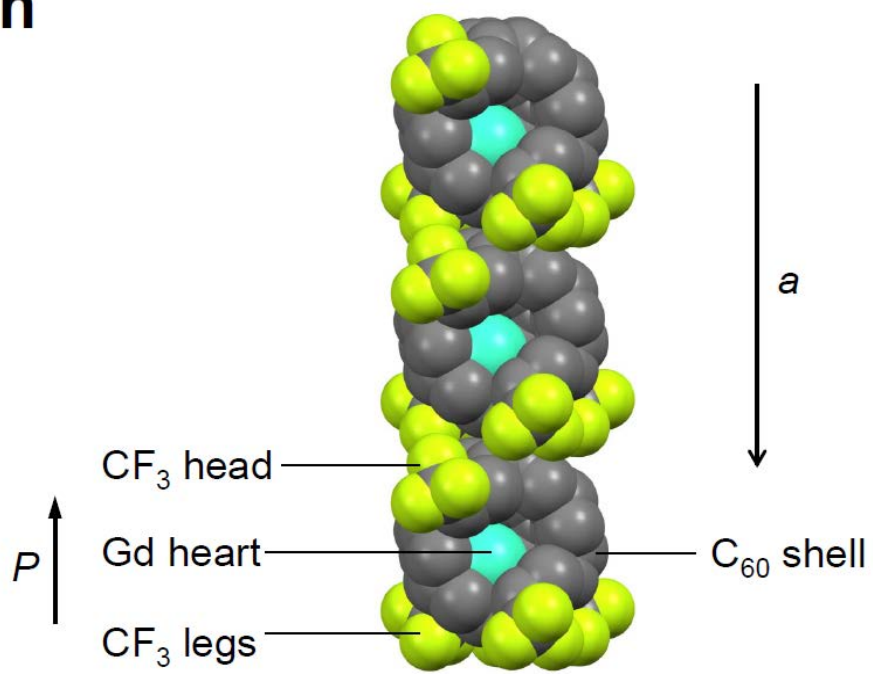

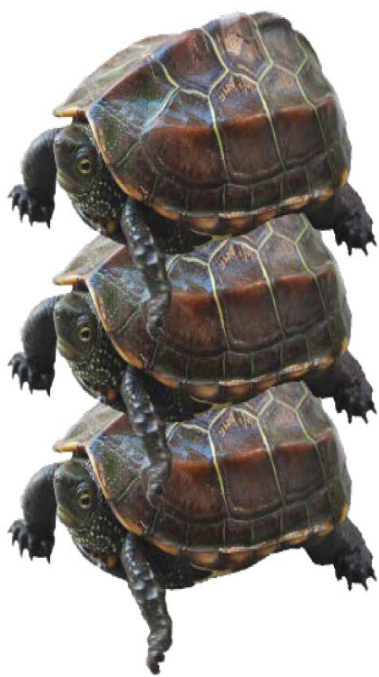

**Fig. 1**

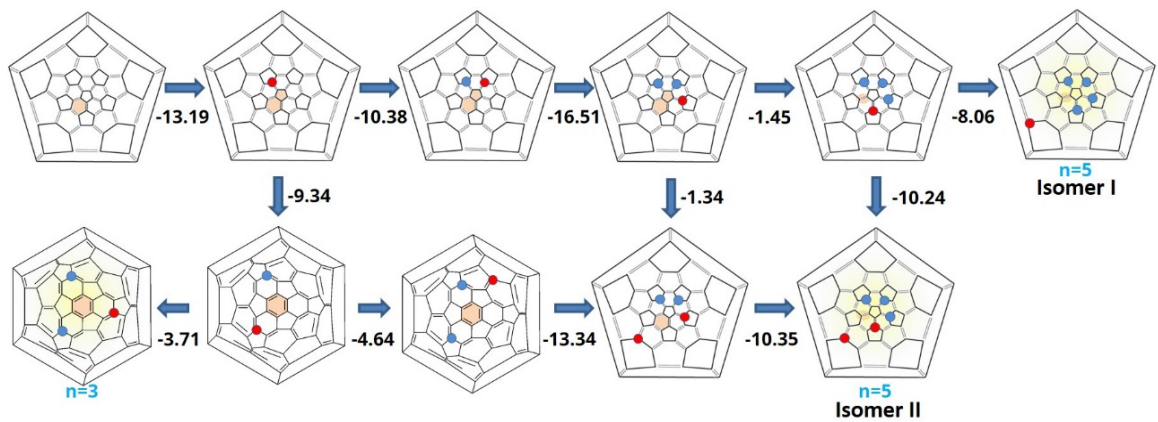

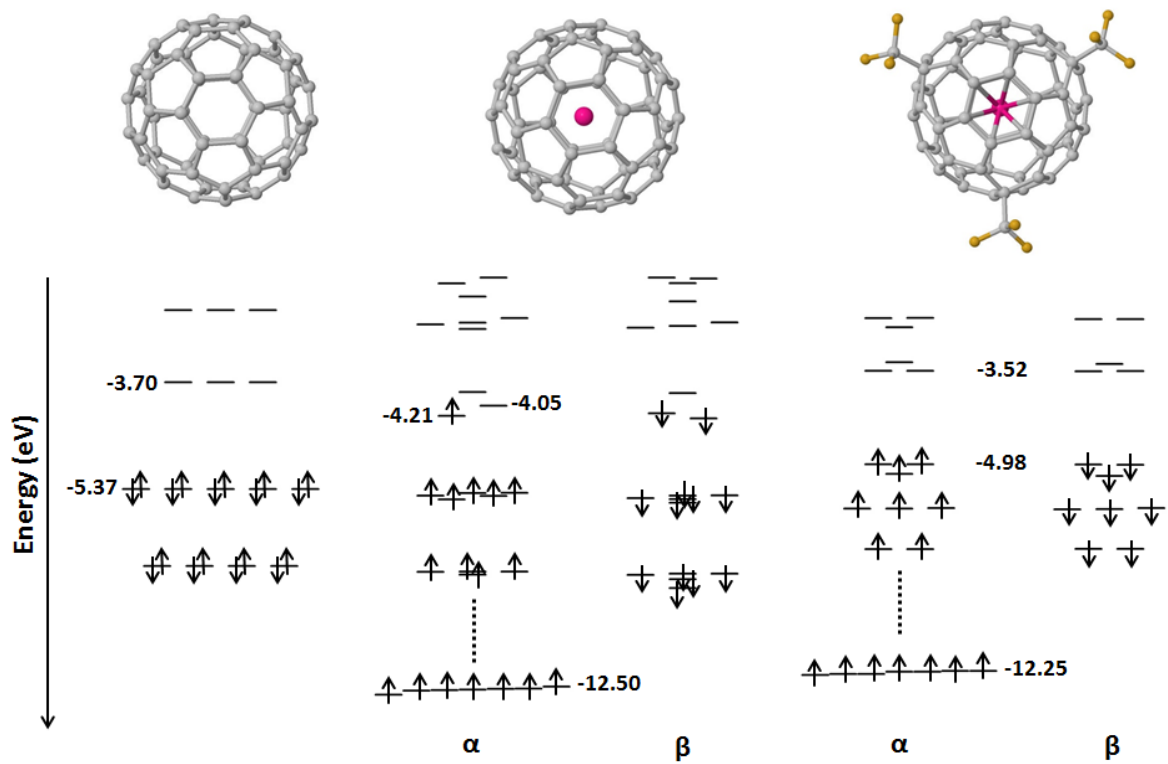

c

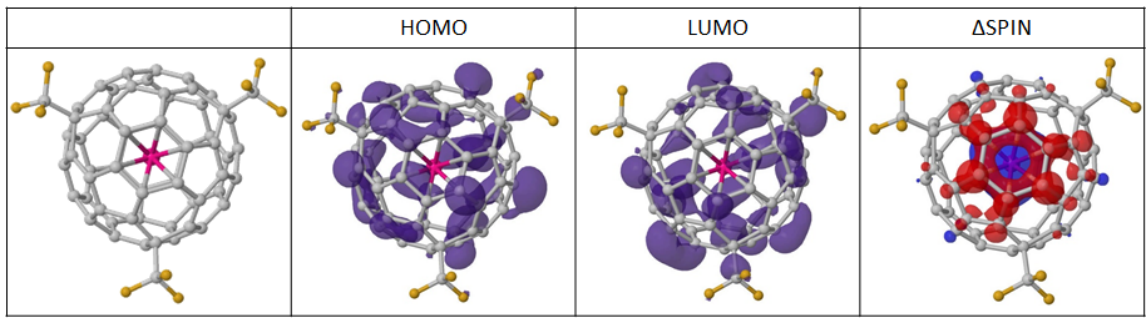

**Fig. 2**

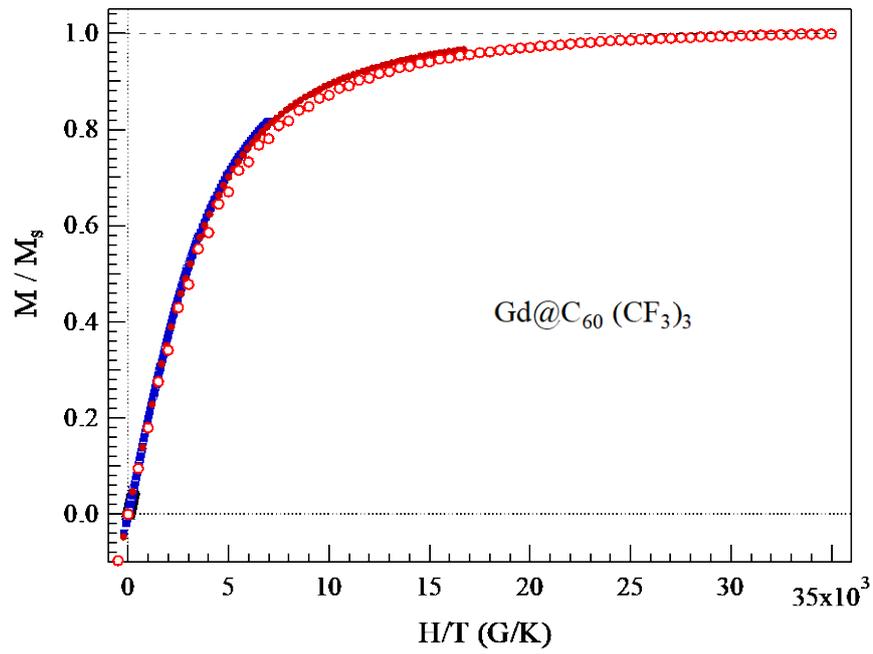

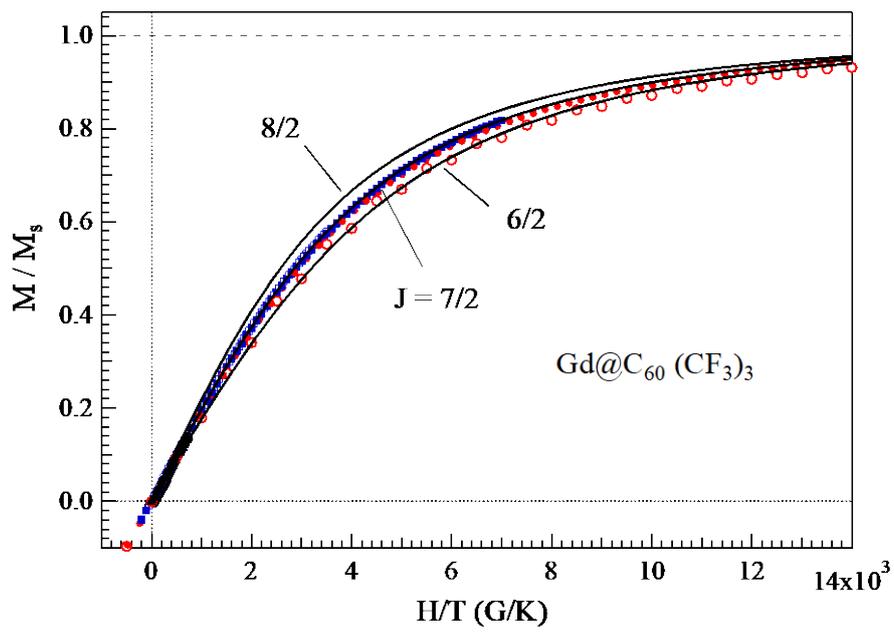

**Fig. 3**

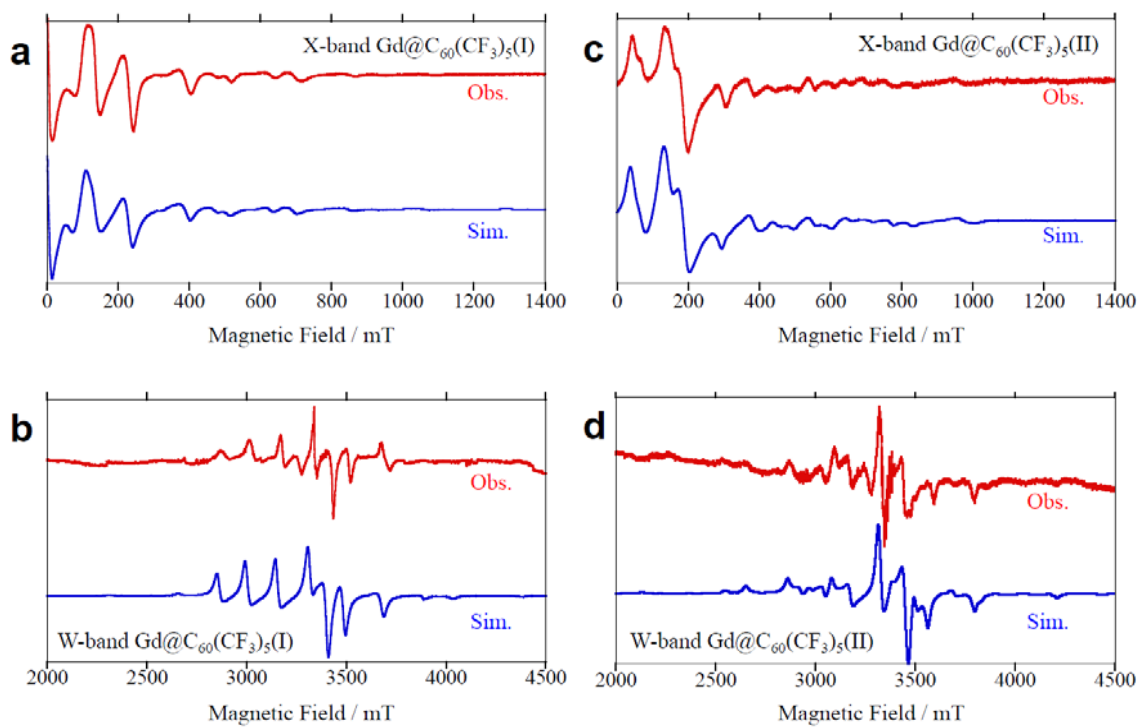

**Fig. 4**